\documentclass[aps,prb,twocolumn,showpacs,superscriptaddress]{revtex4-1}

\usepackage{pdfpages}
\usepackage{amsmath}
\usepackage{amssymb}
\usepackage{amsfonts}
\usepackage{graphicx,graphics}

\newcommand{\abs}[1]{\lvert #1 \rvert}
\newcommand{\expect}[1]{\langle #1\rangle}

\begin{document}

\title{Charge-carrier-induced frequency renormalization, damping and heating of vibrational modes in
  nanoscale junctions}

\author{Kristen Kaasbjerg}
\email{cosby@fys.ku.dk}
\affiliation{School of  Chemistry, The Sackler Faculty of Exact Sciences, 
  Tel Aviv University, Tel Aviv 69978, Israel} 
\author{Tom{\'a}{\v s} Novotn{\'y}}
\affiliation{Department of Condensed Matter Physics, Faculty of Mathematics and
  Physics, Charles Univeristy in Prague, Ke Karlovu 5, 12116 Prague, Czech Republic}
\author{Abraham Nitzan}
\affiliation{School of  Chemistry, The Sackler Faculty of Exact Sciences, 
  Tel Aviv University, Tel Aviv 69978, Israel} 

\date{\today}

\begin{abstract}
  In nanoscale junctions the interaction between charge carriers and the local
  vibrations results in renormalization, damping and heating of the vibrational
  modes. We here formulate a nonequilibrium Green's functions based theory to
  describe such effects. Studying a generic junction model with an off-resonant
  electronic level, we find a strong bias dependence of the frequency
  renormalization and vibrational damping accompanied by pronounced nonlinear
  vibrational heating in junctions with intermediate values of the coupling to
  the leads. Combining our theory with {\em ab-initio} calculations we
  furthermore show that the bias dependence of the Raman shifts and linewidths
  observed experimentally in an OPV3 junction [D. Ward \textit{et al.}, Nature
  Nano.~\textbf{6}, 33 (2011)] may be explained by a combination of dynamic
  carrier screening and molecular charging.
\end{abstract}

\maketitle

\emph{Introduction.}---A fundamental understanding of the nonequilibrium
behavior of the atomic degrees of freedom in current-carrying nanoscale
junctions is of major importance for the realization of molecule-based
electronics. Joule heating arising from the interaction between the electronic
charge carriers and the molecular vibrations, has been observed
experimentally~\cite{Tao:Local,Pascual:ResonantHeating,Selzer:Detection,Natelson:Heating}
and poses a serious stability issue in such junctions.

The implications of the electron-vibration (el-vib) interaction on both the
current and heating in molecular junctions have been subject of intense
theoretical studies~\cite{Nitzan:Vibrational} and critical effects such as
vibrational
instabilities~\cite{Thoss:VibrationalInstabilities,Brandbyge:Laserlike} and
cooling mechanisms~\cite{Nitzan:Cooling,Carlo:Heating,Thoss:Resonant} have been
addressed. In addition, recent
developments~\cite{Todorov:CurrentDriven,Hedegaard:Blowing,Oppen:CurrentInduced,Dundas:CurrentInduced}
have demonstrated the existence of nonconservative current-induced forces that
can drive the molecular vibrations strongly out of equilibrium and provide
further channels for junction destabilization. Efforts to identify vibrational
heating and instabilities in, e.g., inelastic-tunneling spectroscopy
(IETS)~\cite{Brandbyge:Modeling,*Frederiksen:Inelastic}, current
noise~\cite{Belzig:NoiseIII,Aharony:FullCounting} and Raman
spectroscopy~\cite{Nitzan:RamanFromNonequilibrium,Nitzan:RamanInCurrent,Dundas:CurrentInduced}
are still ongoing.

Raman spectroscopy on current-carrying molecular
junctions~\cite{Selzer:Detection,Natelson:Heating} offers a unique diagnostic
tool for monitoring the nonequilibrium behavior of the molecular
vibrations. Apart from the information about heating encoded in the
Stokes/anti-Stokes ratio, Raman spectroscopy provides direct access to the
vibrational spectral function of the molecule. Experimentally, this has revealed
noticeable frequency shifts and broadenings of the vibrations as a function of
bias voltage~\cite{Natelson:Heating}. Similar effects originating from the
electron-phonon interaction are well known from, e.g., Raman spectroscopy on
gated graphene~\cite{Mauri:Breakdown,Pinczuk:Tuning} and current-carrying
nanomechanical carbon-nanotube
resonators~\cite{Zant:StrongCoupling,Bachtold:Coupling,Steele:Probing,Wernsdorfer:DynamicsAndDissipation},
and provide useful insight into the mechanisms governing vibrational damping and
heat dissipation in such systems.

\begin{figure}[!b]
  \centering
  \includegraphics[width=0.95\linewidth]{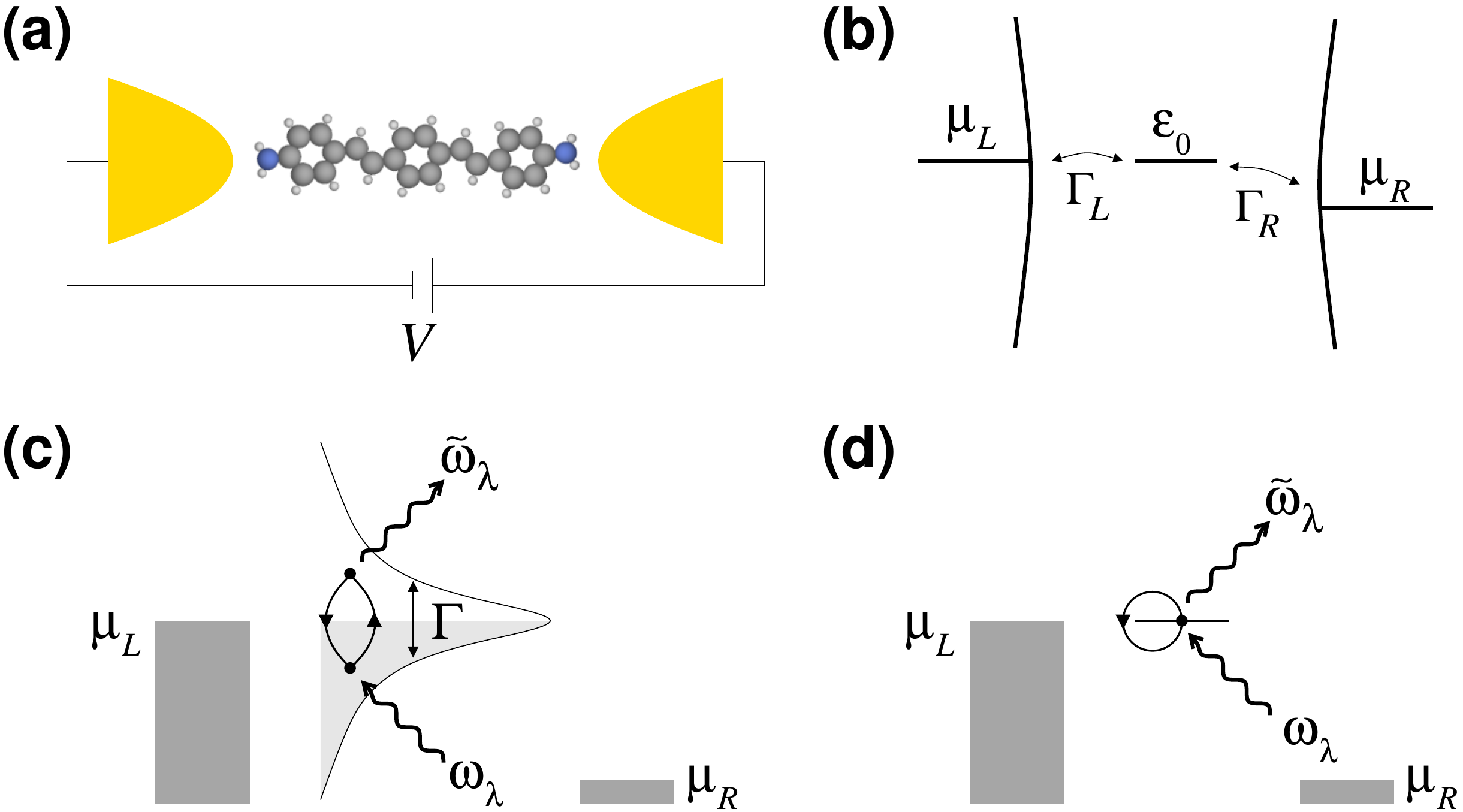}
  \caption{(Color online) (a) Molecular contact with an amine-terminated
    (NH$_2$) OPV3 molecule sandwiched between the source and drain
    electrodes. (b) Generic junction description in terms of a single molecular
    level $\varepsilon_0$ with lead-induced broadening $\Gamma = \Gamma_L +
    \Gamma_R$ and applied bias voltage $eV=\mu_R - \mu_L$. (c),(d) Processes
    responsible for the carrier-induced renormalization of vibrational
    frequencies ($\omega_\lambda\rightarrow\tilde{\omega}_\lambda$) and
    electron-hole damping ($\gamma_{\lambda}^\text{eh}$). The illustrations
    correspond to the Feynman diagrams for the self-energies in
    Eqs.~\eqref{eq:Pi1} and~\eqref{eq:Pi2}, respectively, describing the dynamic
    renormalization via \emph{virtual} electron-hole pairs (screening) (c) and
    the static renormalization due to (partial) charging of the level (d).}
\label{fig:overview}
\end{figure}
In this work we study these effects in nanoscale atomic/molecular junctions
where nonadiabatic (dynamic) effects become important when the electronic and
vibrational energy scales are comparable. For this purpose, we formulate a
nonequilibrium Green's function (NEGF) based theory for vibrational frequency
renormalization and heating taking into account electronic charging and
screening effects on the local junction vibrations. We find that (i) strong
bias-dependent frequency renormalization and damping is accompanied by
pronounced nonlinear bias-dependent heating, and (ii) the above-mentioned
carrier-related mechanisms may explain the voltage dependence of the Raman
shifts and linewidths observed in Ref.~\onlinecite{Natelson:Heating}.

\emph{Theory.}---To address the effect of carrier-induced renormalization and
heating of vibrations in biased nanoscale junctions, we employ the
NEGF~\cite{Jauho} which treats the dynamics of the electronic and vibrational
degrees of freedom on equal
footing~\cite{Hafner:ElvibGoldWire,Galperin:PeaksDips}. In the general case with
more than one local junction vibration, the dressed vibrational retarded Green's
function (GF) is given in the matrix form as
\begin{equation}
  \label{eq:phonon_gf}
  \mathbf{D}^r (\omega) = 
  \left[ \left[\mathbf{D}_0^r(\omega)\right]^{-1} - \mathbf{\Pi}^r(\omega) \right]^{-1} ,
\end{equation}
where $\mathbf{D}_0^r(\omega)$ is the bare diagonal GF given by
$D_{0,\lambda}^r(\omega)=2\omega_\lambda/[(\omega + i0^+)^2 - \omega_\lambda^2]$
for the vibrational mode $\lambda$ with frequency $\omega_\lambda$, and the
spectral function is given by the imaginary part of the retarded GF as
$A(\omega)= -2 \sum_\lambda \mathrm{Im} D_{\lambda\lambda}^r(\omega)$. The
vibrational self-energy $\Pi = \Pi^{\text{el}} + \Pi^{\text{ph}}$ in general has
contributions from both the interaction with electronic charge carriers
($\Pi^{\text{el}}$) and phonon-related effects
($\Pi^{\text{ph}}$)~\cite{footnote5}. The latter, reflecting coupling to
environmental phonons as well as
anharmonicity~\cite{Mingo:Anharmonic,Carlo:HeatDissipation,Jauho:DampingGoldChains,Marzari:Wannier}
is here described by a phenomenological damping rate
$\gamma_{\text{ph}}$~\cite{Galperin:PeaksDips} which does not affect the
vibrational frequencies.

For the electronic part of the self-energy, we consider a junction where the
transport is dominated by a single electronic level (HOMO or LUMO) with energy
$\varepsilon_0$ and level broadening $\Gamma=\Gamma_L + \Gamma_R$ due to the
leads (see Fig.~\ref{fig:overview}(b)). The level is coupled to the local
vibrations via linear $H^{(1)} = n_0 \sum_\lambda M_\lambda^{(1)}
(a_\lambda^\dagger + a_\lambda^{\phantom\dagger})$ and quadratic $H^{(2)} =
\frac{1}{2} n_0 \sum_{\lambda\lambda'} (a^\dagger_\lambda +
a^{\phantom\dagger}_\lambda) M_{\lambda\lambda'}^{(2)} (a^\dagger_{\lambda '} +
a^{\phantom\dagger}_{\lambda '})$ el-vib interaction terms where $n_0=n_\uparrow
+ n_\downarrow$ is the level occupancy and $M_\lambda^{(1)}$
($M_{\lambda\lambda'}^{(2)}$) the linear (quadratic) el-vib coupling. The
retarded components of the electronic self-energy $\Pi^\text{el} = \Pi^{(1)} +
\Pi^{(2)}$, are to the lowest (nonzero) order given by the electronic
polarizability~\cite{Galperin:PeaksDips}
\begin{align}
  \label{eq:Pi1}
  \Pi_{\lambda\lambda'}^{(1), r}(\omega) & = -2 i \int \! 
  \frac{d\varepsilon}{2\pi} \; 
  \big[ M_\lambda^{(1)} G^r(\varepsilon) 
        M_{\lambda'}^{(1)} G^<(\varepsilon - \omega) \nonumber \\
    & \quad + M_\lambda^{(1)} G^<(\varepsilon) 
              M_{\lambda'}^{(1)} G^a(\varepsilon - \omega)
  \big]   ,
\end{align}
and the level occupancy~\cite{supplemental}
\begin{equation}
  \label{eq:Pi2}
  \Pi_{\lambda\lambda'}^{(2), r}(\omega) = - 2 i M_{\lambda\lambda'}^{(2)} \int 
     \! \frac{d\varepsilon}{2\pi} \; 
     G^<(\varepsilon) = \expect{n_0} M_{\lambda\lambda'}^{(2)} ,
\end{equation}
respectively, where the factors of 2 account for spin degeneracy. The two
self-energies, which are illustrated in Fig.~\ref{fig:overview}(c) and~(d) by
their respective Feynman diagrams, account for frequency renormalization due to
dynamic screening and static (partial) charging, respectively. In addition, the
former also accounts for spectral broadening due to damping by electron-hole
(eh) pair excitations.

Vibrational heating originates from the linear el-vib interaction and can be
quantified in terms of the nonequilibrium occupation $\tilde{n}_\lambda$ of the
renormalized vibration. As we here show in detail~\cite{supplemental}, the
physically relevant (i.e., experimentally verifiable via, e.g., Raman
spectroscopy) occupation factor for a renormalized vibration is given by
\begin{equation}
  \label{eq:occupation}
  \tilde{n}_\lambda = 
    -\frac{1}{2} + \frac{\tilde{\omega}_\lambda}{\omega_\lambda} 
    \frac{i}{2} \int \! \frac{d\omega}{2\pi} D_{\lambda\lambda}^<(\omega)  
\end{equation}
where $\tilde{\omega}_\lambda$ is the renormalized frequency,
$D_{\lambda\lambda}^<$ are the diagonal elements of the vibrational lesser GF
$\mathbf{D}^<(\omega) = \mathbf{D}^r(\omega) \mathbf{\Pi}^<(\omega)
\mathbf{D}^a(\omega)$, and the \emph{mode-renormalization} factor
$\tilde\omega_\lambda/ \omega_\lambda$ accounts for the renormalization of the
field operators $\tilde{a}_\lambda^\dagger + \tilde{a}^{\phantom\dagger}_\lambda
= x_\lambda \sqrt{2\tilde{\omega}_\lambda}$ here related to the vibrational
normal coordinate $x_\lambda$ through the \emph{renormalized} frequency
$\tilde{\omega}_\lambda$. 

\begin{figure}[!t]
  \begin{minipage}{1.0\linewidth}
    \hspace{-2mm}
    \includegraphics[width=0.49\linewidth]{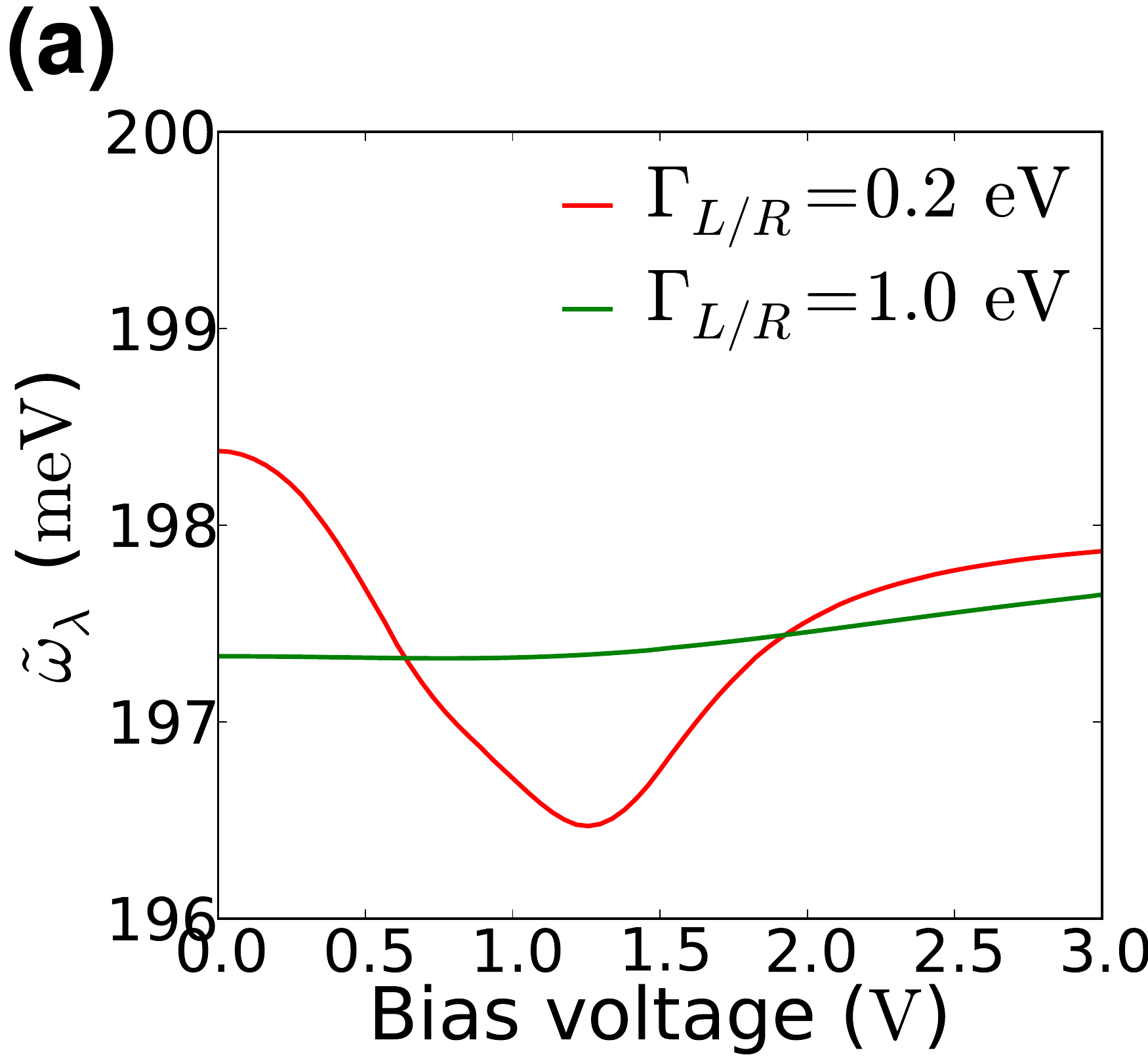}
    \includegraphics[width=0.49\linewidth]{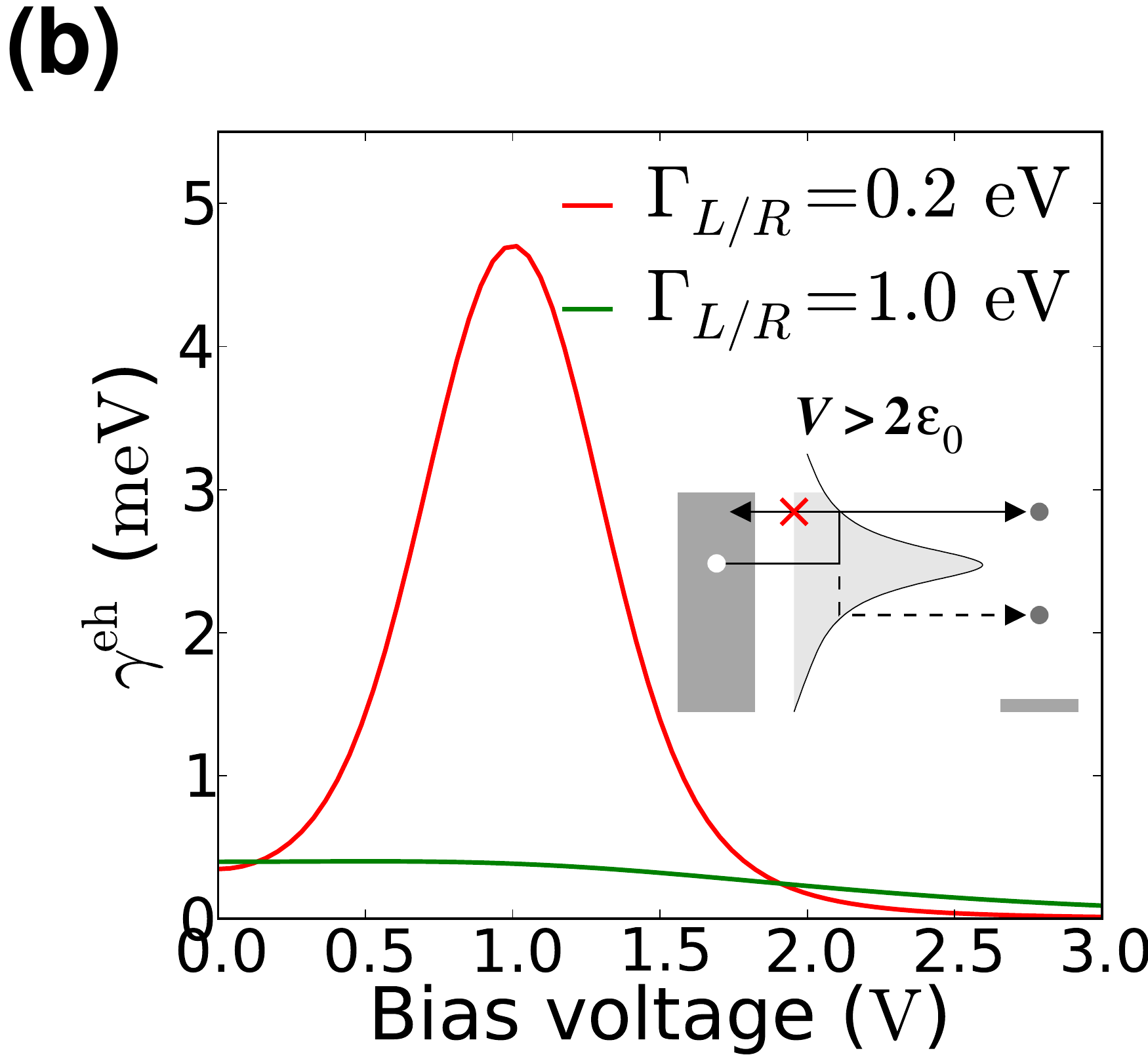}
  \end{minipage}
  \begin{minipage}{1.0\linewidth}
    \hspace{-2mm}
    \includegraphics[width=0.49\linewidth]{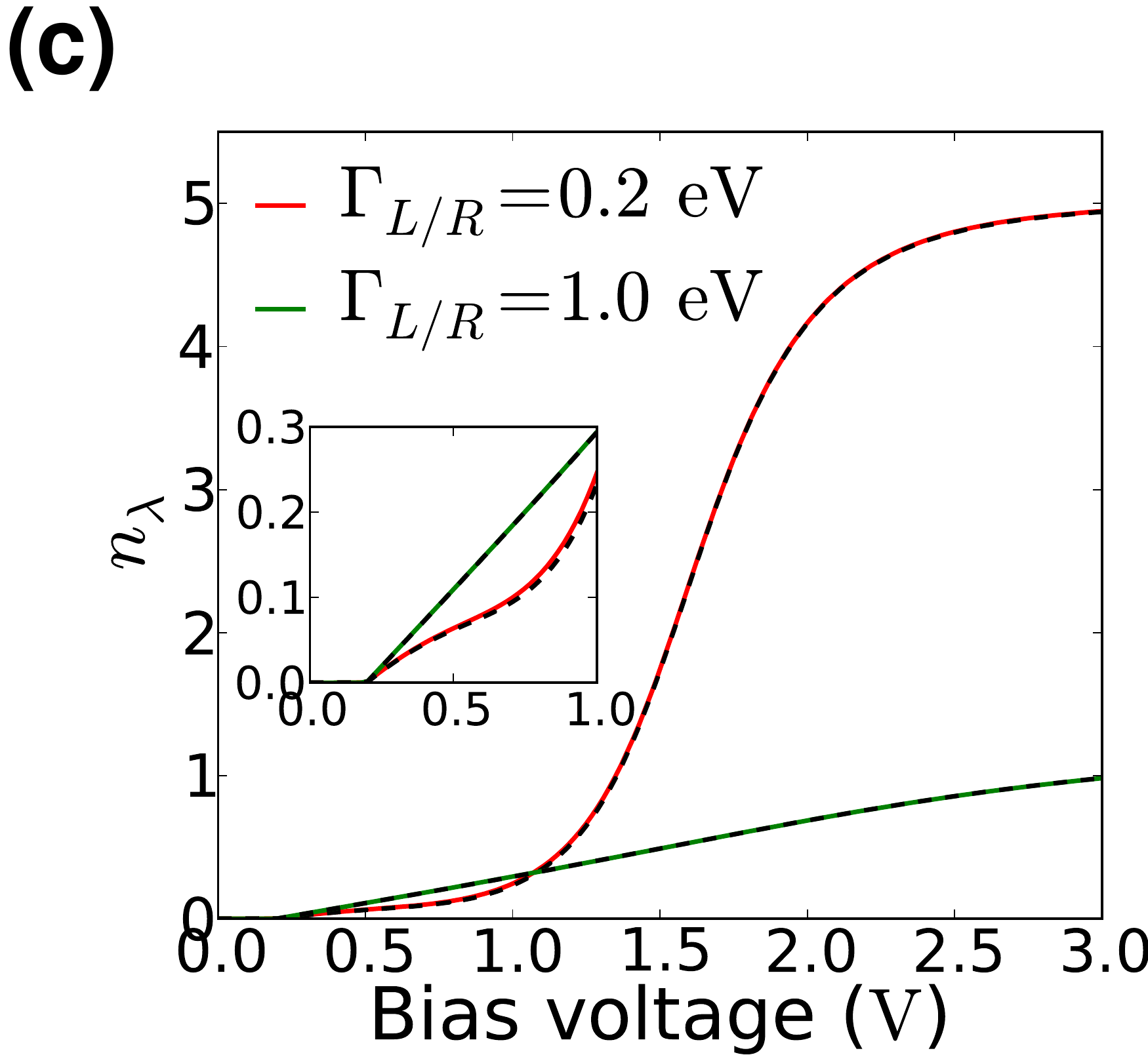}
    \includegraphics[width=0.49\linewidth]{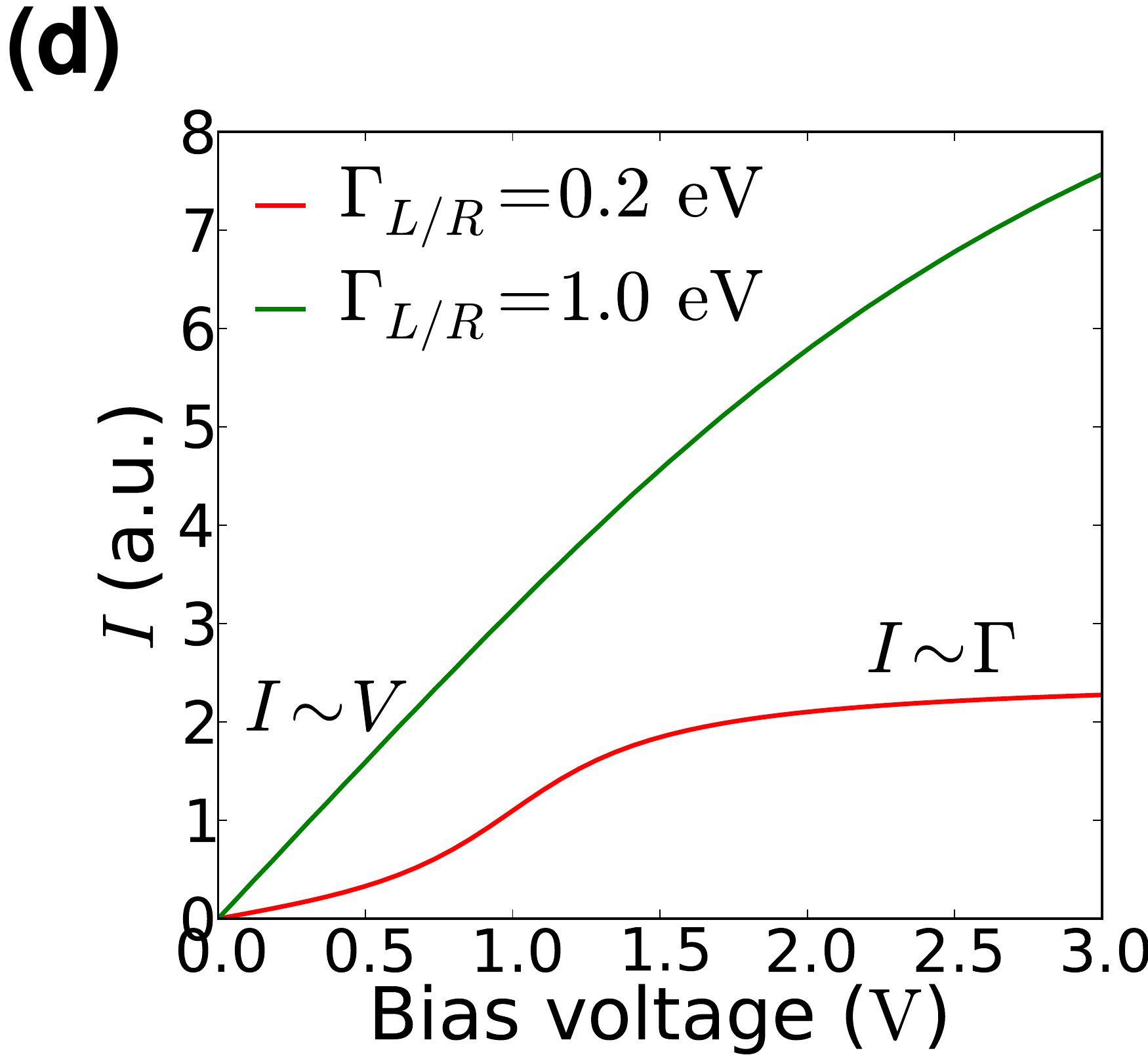}
  \end{minipage}
  \caption{(Color online) Renormalized frequency $\tilde{\omega}_\lambda$ (a)
    and electron-hole damping rate $\gamma^\text{eh} = -2
    \tfrac{\omega_{\lambda}}{\tilde{\omega}_{\lambda}} \mathrm{Im}
    \Pi_\lambda^{(1),r}(\tilde{\omega}_\lambda)$ (b) as a function of bias
    voltage $V$ for a vibration interacting with an off-resonant level
    ($\varepsilon_0=0.5$~eV, $\mu_{L/R} = \pm V/2$, $\Gamma_L=\Gamma_R$,
    $\omega_\lambda=200$~meV, $M_\lambda^{(1)} = 50$~meV, $M_\lambda^{(2)} =
    -2$~meV, $\gamma_\text{ph}=1$~meV). The inset in (b) depicts the eh-pair
    processes dominating the damping rate. (c) Nonequilibrium excitation
    $\tilde{n}_\lambda$ of the renormalized vibration calculated using
    Eq.~\eqref{eq:occupation} (full lines) and the rate-equation result in
    Eq.~\eqref{eq:n_steadystate} (black dashed lines). (d) Electronic current
    through the junction.}
  \label{fig:omega_vs_V}
\end{figure}

The occupation in Eq.~\eqref{eq:occupation} is equivalent to the steady-state
solution of the rate equation~\cite{supplemental}
\begin{align}
  \label{eq:rateequation}
  \dot{\tilde{n}}_\lambda & =
  \left[
    \gamma_\text{emis}^\text{eh} + \gamma_{\text{ph}}
    N_{B}(\tilde{\omega}_\lambda)\right] (\tilde{n}_\lambda + 1) \nonumber \\ 
  & \quad - \left[\gamma_\text{abs}^\text{eh} +
    \gamma_\text{ph}(N_{B}(\tilde{\omega}_\lambda) + 1)
  \right] \tilde{n}_\lambda, 
\end{align}
given by
\begin{equation}
  \label{eq:n_steadystate}
  \tilde{n}_\lambda = 
  \frac{1}{e^{\hbar\tilde{\omega}_\lambda/ k_\text{B} T_\text{eff}} - 1} ,
\end{equation}
where we have introduced an \emph{effective} nonequilibrium mode temperature
$T_\text{eff}$ defined by $e^{\hbar\tilde{\omega}_\lambda/ k_\text{B}
  T_\text{eff}} = [ \gamma_\text{abs}^\text{eh} + \gamma_\text{ph}(N_{B}+1) ]/
[\gamma_\text{emis}^\text{eh}+\gamma_{\text{ph}}N_{B}]$, $N_B(\omega) =
[e^{\hbar\omega / k_\text{B} T} - 1 ]^{-1}$ is the Bose-Einstein occupation
factor of the environmental phonon bath at temperature $T$, and
$\gamma_\text{abs}^\text{eh}$ ($\gamma_\text{emis}^\text{eh}$) is the rate for
absorption (emission) of vibrational quanta due to \emph{intra} and
\emph{inter}electrode eh-pair processes [see inset in
Fig.~\ref{fig:omega_vs_V}(b)]. They can be identified
\cite{Brandbyge:Modeling,*Frederiksen:Inelastic} from the eh-pair damping rate
$\gamma_\lambda^\text{eh} = -2
\tfrac{\omega_{\lambda}}{\tilde{\omega}_{\lambda}}\mathrm{Im}
\Pi_{\lambda\lambda}^{(1),r}(\tilde{\omega}_\lambda)$~\cite{supplemental} as
$\gamma_\lambda^\text{eh}= i \tfrac{\omega_{\lambda}}{\tilde{\omega}_{\lambda}}
( \Pi_{\lambda\lambda}^{(1),>} - \Pi_{\lambda\lambda}^{(1),<} ) =
\gamma_\text{abs}^\text{eh} - \gamma_\text{emis}^\text{eh}$. In equilibrium they
are related by the detailed balance
$\gamma_\text{emis}^\text{eh}=\gamma_\text{abs}^\text{eh} \exp{
  (-\tfrac{\hbar\tilde{\omega}_\lambda}{k_\text{B}T}) }$ implying
$T_{\text{eff}}=T$. The total damping rate is given by the imaginary part of the
full self-energy as $\gamma_\lambda = -2 \tfrac{\omega_{\lambda}}
{\tilde{\omega}_{\lambda}}\mathrm{Im} \Pi_{\lambda\lambda}^{r}
(\tilde{\omega}_\lambda)$.

In the case of independent vibrations, i.e., when mode-mode couplings given
by the off-diagonal elements of the self-energy are negligible, the renormalized
frequencies are given by the real part of the self-energy as the solution to the
equation
\begin{equation}
  \label{eq:omega_renormalized}
  \omega^2 = \omega_\lambda^2 + 2\omega_\lambda \text{Re}
  \Pi_{\lambda\lambda}^r(\omega) ,
\end{equation}
which for small frequency changes $\Delta\omega_\lambda \ll \omega_\lambda$
simplifies to $\Delta\omega_\lambda = \text{Re}
\Pi_{\lambda\lambda}^r(\omega_\lambda)$. In the limit $\Gamma \gg
\omega_\lambda,\varepsilon_0, V$ corresponding to the situation in atomic gold
wires~\cite{Jauho:Inelastic,Brandbyge:Modeling,Hafner:ElvibGoldWire,Frederiksen:Inelastic},
the renormalization and damping due to linear el-vib interaction
[Eq.~\eqref{eq:Pi1}] scale as $\Delta\omega_\lambda^{(1)} \sim -
(M_\lambda^{(1)})^2 / \Gamma = -\omega_\lambda(M_\lambda^{(1)} /
\Gamma)^2\Gamma/\omega_\lambda$ and $\gamma^\text{eh}_\lambda \sim
\omega_\lambda (M_\lambda^{(1)} / \Gamma)^2$, respectively. Even if the
dimensionless coupling constant is small $(M_\lambda^{(1)} / \Gamma)^2\ll 1$,
the frequency renormalization may become appreciable if
$\Gamma\gg\omega_{\lambda}$. In this regime, the omission of the
\emph{mode-renormalization} in Eq.~\eqref{eq:occupation} has previously led to
disagreements between NEGF and rate equation results for the vibrational
occupation (see, e.g., Fig.~2(a) of Ref.~\onlinecite{Yeyati:Nonlinear}).

In the calculations presented below, we assume low temperature ($T=0$) and use
the \emph{bare} electronic GFs in the evaluation of the self-energies
corresponding to the first Born approximation.

\emph{Generic junction model.}---In order to demonstrate the connection between
vibrational frequency renormalization, broadening/damping, and heating in
nanoscale junctions, we start by considering a simple junction with an
off-resonant electronic level $\varepsilon_0$ coupled symmetrically to the leads
($\Gamma_L=\Gamma_R$) and interacting weakly with a local vibration with
frequency $\omega_\lambda$. The significance of the electronic energy scales of
the junction is illustrated by considering two cases for the lead-induced level
broadening representing junctions with $\Gamma\lesssim \varepsilon_0, V$ and
$\Gamma\gg\varepsilon_0, V$, respectively (see Fig.~\ref{fig:omega_vs_V} for
parameters).

In Fig.~\ref{fig:omega_vs_V}(a) and~(b) we plot the bias dependence of the
renormalized vibrational frequency $\tilde{\omega}_\lambda$ obtained from a
self-consistent solution of Eq.~\eqref{eq:omega_renormalized} and the eh-pair
damping rate $\gamma^\text{eh}$, respectively. The qualitatively different bias
dependencies in the two cases stem from different electronic density of states
(DOS) of the level. For the largest value of $\Gamma_{L/R}$ in
Fig.~\ref{fig:omega_vs_V} corresponding to, e.g., break junctions with small
molecules~\cite{Ruitenbeek:Stretching}, the DOS is low and constant at the scale
of the applied bias. The damping is consequently weak ($\gamma^\text{eh}\lesssim
0.4$~meV) and both the frequency renormalization and damping show very little
bias dependence.

When $\Gamma\lesssim \varepsilon_0, V$, the electronic resonance can be
introduced into the bias window resulting in large changes in the level
occupancy and the electronic screening with the applied bias. This situation
initially results in a softening of the vibration and increased damping with
applied bias. At $V=2\varepsilon_0$ where the level becomes resonant with the
chemical potential of the source contact, it is partially filled and therefore
has a large DOS for eh-pair excitations [see Fig.~\ref{fig:overview}(c)]. As a
consequence, the frequency renormalization and eh-pair damping peak close to
this bias voltage value. For higher biases the resonance continues to fill up
and eh-pair excitations become suppressed by Pauli blocking [see inset of
Fig.~\ref{fig:omega_vs_V}(b)]. This results in a subsequent hardening of the
frequency, which at high bias saturates at the value given by the
charging-induced renormalization $\tilde{\omega}_\lambda = \omega_\lambda + n_0
M_\lambda^{(2)}$, and a strongly reduced damping (see also below).

Next, we consider the heating of the
vibration~\cite{footnote4}\nocite{Wang:Coupled}. Figure~\ref{fig:omega_vs_V}(c)
and~(d) show the bias dependence of the vibrational excitation obtained from
Eq.~\eqref{eq:occupation} as well as the rate equation
result~\eqref{eq:n_steadystate}, and the elastic current $I=-e \Gamma/h \int \!
d\varepsilon \, \mathrm{Im}G^r(\varepsilon) \left[f_L - f_R \right] $ through
the junction (inelastic contributions are small corrections). Note that the
rate-equation and NEGF results are in perfect agreement as expected. Above the
threshold for emission of vibrational quanta at $eV=\hbar\tilde{\omega}_\lambda$
(given by the renormalized frequency) a qualitatively different behavior of the
vibrational excitation is observed for the two $\Gamma$ values. In the
$\Gamma\gg \varepsilon_0, V$ case, the linear bias dependence of the vibrational
excitation is well-known~\cite{Frederiksen:Inelastic}. For $\Gamma\lesssim
\varepsilon_0, V$, the bias dependence of the vibrational excitation is strongly
nonlinear and, perhaps counter intuitively, experiences stronger (compared to
the large $\Gamma$ case) heating for $V\gtrsim 1.1$~V despite the lower current
and stronger damping.

To elucidate the physical origin of the pronounced nonlinear heating in the
$\Gamma\lesssim \varepsilon_0, V$ case, we consider the dominant contributions
to the eh-pair damping rate from absorption (full lines) and emission (dashed
line) processes shown in the inset of Fig.~\ref{fig:omega_vs_V}(b). At biases
$V> 2\varepsilon_0$, absorption processes with the left lead become
Pauli-blocked and transport-induced \emph{inter}electrode processes become
dominant. Furthermore, when the full resonance is contained in the bias window,
i.e., $V \gg \varepsilon_0,\Gamma, \omega_\lambda$, the DOS for absorption and
emission processes become comparable implying that $\gamma_\text{abs}^\text{eh}
\sim \gamma_\text{emis}^\text{eh}$ and $\gamma^\text{eh} \rightarrow 0$. For the
steady-state solution~\eqref{eq:n_steadystate} this leads ($N_{B}=0$ at $T=0$)
to $\tilde{n}_\lambda = \gamma_\text{emis}^\text{eh} / \gamma_\text{ph}$. The
bias dependence of the heating thus follows that of the rate for emission
processes which is nonlinear due to the Lorentzian broadening of the electronic
level and saturates in the high bias limit where the current is carried by the
full resonance. This qualitatively explains the pronounced nonlinear heating. It
is important to note that the steady-state solution to the rate
equation~\eqref{eq:rateequation} diverges in the high bias limit $V\to\infty$ in
the absence of the phonon-related damping parameter $\gamma_\text{ph}$. This
underlines the importance of the $\gamma_\text{ph}$ parameter for vibrational
heating and
instabilities~\cite{Thoss:VibrationalInstabilities,Brandbyge:Laserlike,Brandbyge:Nanoconstriction}
in nanoscale junctions.

\emph{OPV3 junction.}--- In the remaining part of the paper, we present
first-principles based calculations of the carrier-induced frequency
renormalization and vibrational heating in a junction based on the
amine-terminated OPV3 molecule [see Fig.~\ref{fig:overview}(a)] where frequency
shifts of the order of $\sim$1~meV, spectral broadening, and significant heating
have been observed experimentally with increasing bias
voltage~\cite{Natelson:Heating}. The frequency shifts were so far
ascribed~\cite{Natelson:Heating} to a vibrational Stark
effect~\cite{Lambert:StarkEffect}. However, calculations of ours of the
vibrational frequencies in the presence of an electric field do not support this
interpretation~\cite{supplemental}.

In our model of the OPV3 junction, transport through the LUMO of the OPV3
molecule positioned off-resonant ($\varepsilon_0=0.5$~eV) with respect to the
equilibrium chemical potentials of the leads ($\mu_{L/R} = \pm V/2$) and coupled
symmetrically to the contacts ($\Gamma_{L/R}=0.2$~eV) is
assumed~\cite{footnote6}. The next molecular level lies $\sim$1~eV above the
LUMO and can hence be neglected. The vibrations and el-vib interactions have
been obtained from first principles for the isolated OPV3
molecule~\cite{supplemental}, thus neglecting direct effects from the
leads. With few exceptions, the quadratic couplings, which are of the order of
$\abs{M_{\lambda\lambda'}^{(2)}}\sim0$--$10$~meV, are found to be negative
corresponding to frequency softening upon charging of the LUMO. This is in good
agreement with a recent study of charging-induced frequency shifts in molecular
junctions~\cite{Ratner:Probing}.
\begin{figure}[!t]
  \includegraphics[scale=0.28]{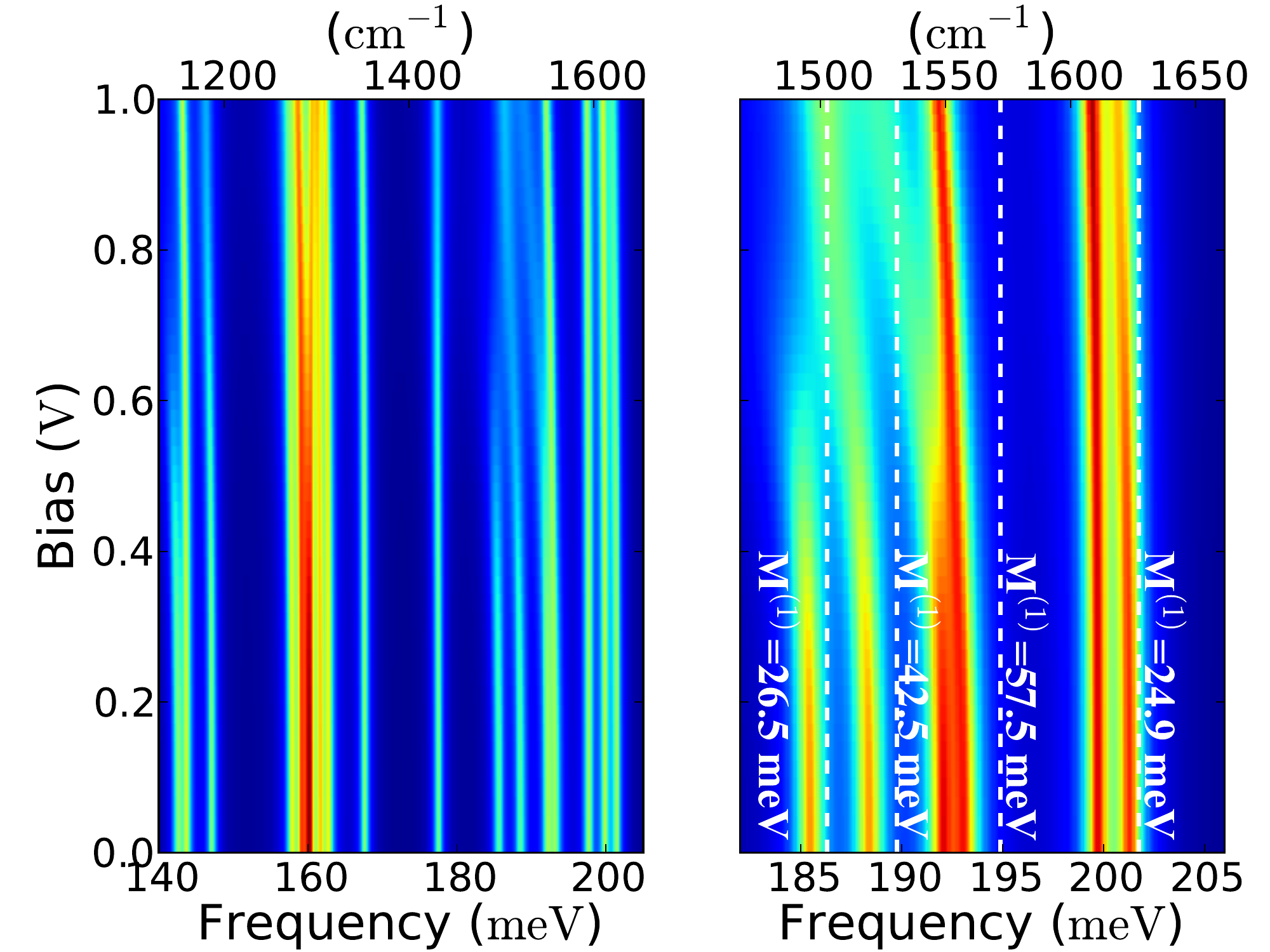}
  \includegraphics[scale=0.28]{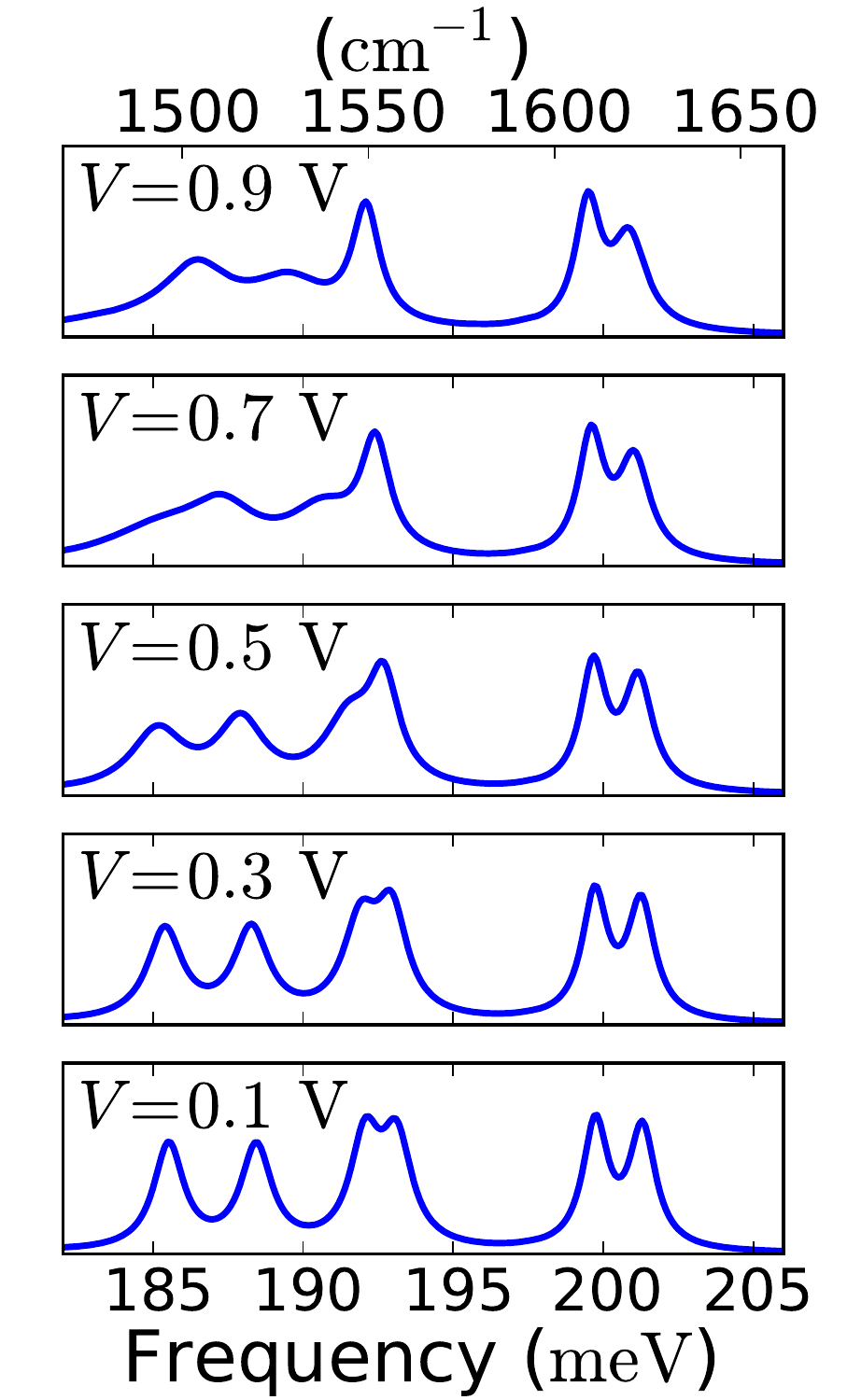}
  \caption{(Color online) Vibrational spectral function $A(\omega)$ for the
    amine-terminated OPV3 molecule as a function of bias voltage. In the
    magnified center plot, the unperturbed frequencies $\omega_\lambda$ of the
    isolated molecule (vertical dashed lines) and the linear el-vib couplings
    $M_\lambda^{(1)}$ are indicated for the modes with strongest el-vib
    interaction. Only the projection of the spectral function onto vibrations
    with significant el-vib coupling ($\abs{M_\lambda^{(1)}} > 5$~meV) is
    shown.}
  \label{fig:opv3NH2_spectral}
\end{figure}

Figure~\ref{fig:opv3NH2_spectral} shows the bias dependence of the vibrational
spectral function $A(\omega)$ for the OPV3 junction. We focus here on the
high-energy vibrations ($\omega_\lambda > 140$~meV), but note that the
low-energy part of the spectral function shows a similar bias
dependence~\cite{supplemental}. Overall, the calculated spectral function
reproduces~\cite{footnote3} the features in the Raman spectra of Fig.~3(b) of
Ref.~\onlinecite{Natelson:Heating}. The spectral lines show clear mode
\emph{softening} of up to $\sim$2~meV with increasing bias voltage (for HOMO
dominated transport, however, many of the spectral lines show mode
\emph{hardening}~\cite{supplemental}). This is a result of partial charging and
screening that follows as the chemical potential of the left lead moves into the
broadened LUMO resonance. The relative magnitude of the two effects is sensitive
to the level alignment and lead broadening in the junction. Here, they
contribute comparably to the frequency renormalization. The eh-pair damping that
accompanies screening, gives rise to pronounced broadening of some of the
spectral lines. This effect correlates with the strength of the linear el-vib
interaction $M_\lambda^{(1)}$ which is indicated in the center panel of
Fig.~\ref{fig:opv3NH2_spectral} for the modes with the strongest interaction. At
zero bias, the spectral peaks for these modes are shifted from the frequencies
of the isolated molecule (vertical dashed lines) due to (equilibrium) charging
and screening. At large bias voltages, el-vib mediated mode-mode couplings
result in a nontrivial bias dependence of some of the closely lying spectral
lines.

\begin{figure}[!b]
  \includegraphics[width=0.65\linewidth]{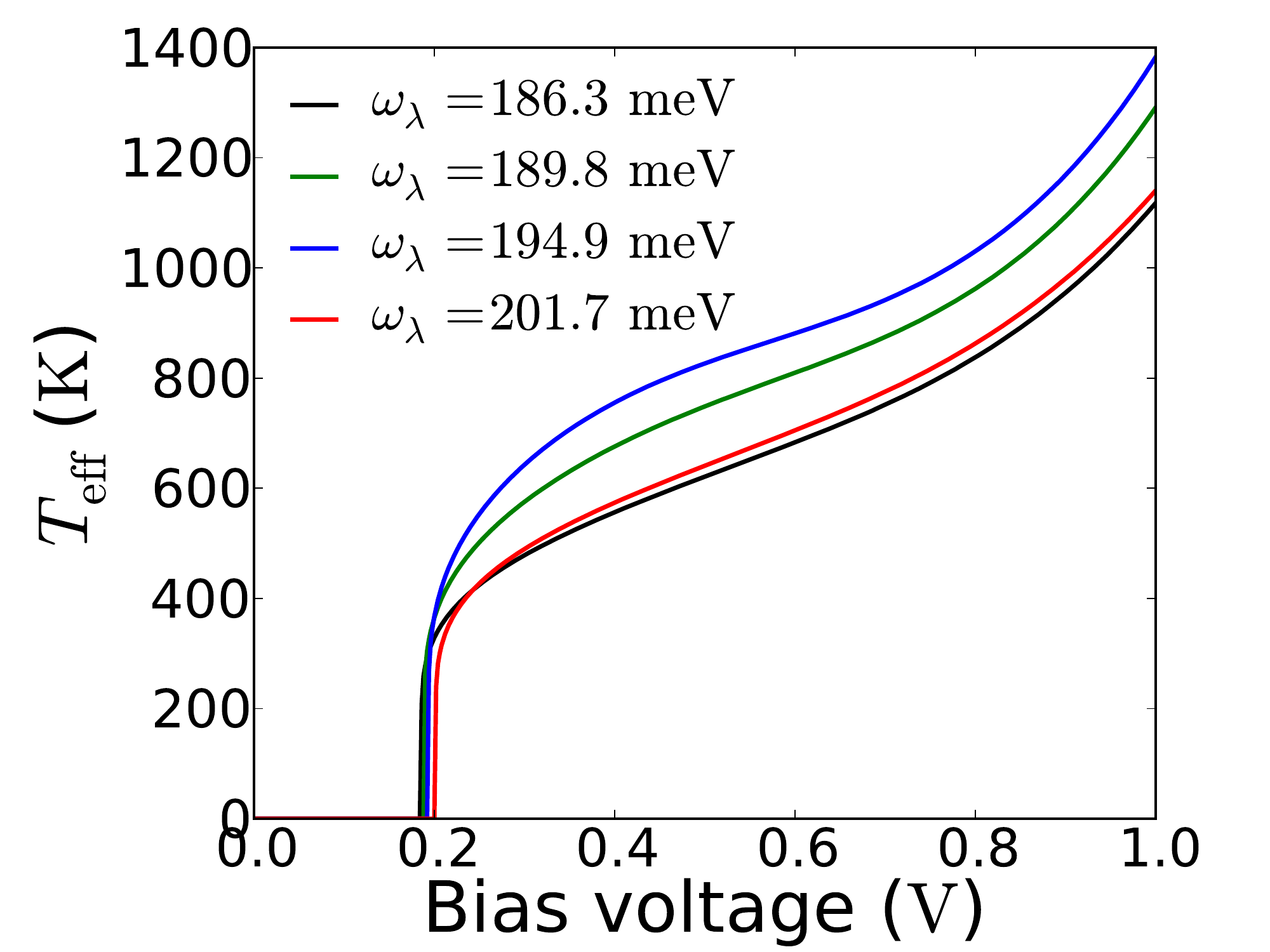}
  \caption{(Color online) Effective nonequilibrium temperature $T_\text{eff}$
    from Eq.~\eqref{eq:n_steadystate} ($\gamma_\text{ph}=1$~meV) for the OPV3
    modes marked with dashed lines in the center panel of
    Fig.~\ref{fig:opv3NH2_spectral}.}
  \label{fig:opv3NH2_temperature}
\end{figure}

In Fig.~\ref{fig:opv3NH2_temperature} we show the \emph{effective}
nonequilibrium temperature $T_\text{eff}$ from Eq.~\eqref{eq:n_steadystate} for
the OPV3 modes marked with dashed lines in the center panel of
Fig.~\ref{fig:opv3NH2_spectral}. Above the emission threshold at
$eV=\hbar\tilde{\omega}_\lambda$, the temperature of the vibrations jumps to
several hundred Kelvin. For the $\omega_\lambda=201.7$~meV mode, the temperature
is in good agreement with the one for the mode with similar energy reported in
Fig.~3(a) of Ref.~\onlinecite{Natelson:Heating}. Comparing to
Fig.~\ref{fig:omega_vs_V}(c), the approximate linear increase in the effective
temperatures above the emission threshold is seen to correspond to the initial
nonlinear increase in the occupation.

\emph{Conclusions.}--- In summary, we have studied carrier-induced vibrational
frequency renormalization, damping, and heating originating from the microscopic
eh-pair excitation processes governing these observables. In junctions
characterized by $\Gamma \lesssim \varepsilon_0, V$ where the effects are most
pronounced, a strong correlation with nonlinear vibrational heating and the
onset of current results (see Fig.~\ref{fig:omega_vs_V}). Such heating is
inherent to junctions where the current is carried by a full electronic
resonance and underlines the importance of eh-pair processes for the damping of
vibrational heating. We have further shown that the voltage dependence of the
Raman shifts and linewidths observed for an OPV3 molecular junction in
Ref.~\onlinecite{Natelson:Heating} is not consistent with the originally
proposed static Stark shift but can be explained by carrier-induced charging and
screening effects.

\begin{acknowledgments}
  We thank M.~Galperin for fruitful discussions. The research of A.N. is
  supported by the Israel Science Foundation, the Israel-US Binational Science
  Foundation and the European Research Council under the European Union's
  Seventh Framework Program (FP7/2007-2013; ERC grant agreement n$^\circ$
  226628). T.N. acknowledges financial support by the Czech Science Foundation
  via Grant No.~204/12/0897 and K.K. support from the Villum Kann Rasmussen
  Foundation.
\end{acknowledgments}

\appendix
\onecolumngrid
\clearpage
\section*{}
\thispagestyle{empty}


\section*{Supplementary material}

# Supplemental material for "Charge-carrier-induced frequency renormalization, damping and heating of vibrational modes in nanoscale junctions"

Kristen Kaasbjerg,[1] Tomáš Novotný,[2] and Abraham Nitzan[1]

[1]*School of Chemistry, The Sackler Faculty of Exact Sciences, Tel Aviv University, Tel Aviv 69978, Israel*
[2]*Department of Condensed Matter Physics, Faculty of Mathematics and Physics,
Charles Univeristy in Prague, Ke Karlovu 5, 12116 Prague, Czech Republic*

## I. JUNCTION HAMILTONIAN

To fix the notation, we start by specifying the Hamiltonian $H = H_{\text{el}} + H_{\text{vib}} + H_{\text{int}}$ for a junction with an electronic level coupled to left ($L$) and right ($R$) leads ($H_{\text{el}}$) and interacting ($H_{\text{int}}$) with a set of vibrations ($H_{\text{vib}}$) through both linear ($H^{(1)}$) and quadratic ($H^{(2)}$) el-vib interaction, i.e.

$$H_{\text{el}} = \varepsilon_0 d^\dagger d + \sum_{k;\alpha=L,R} \varepsilon_{k\alpha} c_{k\alpha}^\dagger c_{k\alpha} + \sum_{k;\alpha=L,R} (t_{k\alpha} c_{k\alpha}^\dagger d + t_{k\alpha}^* d^\dagger c_{k\alpha}), \tag{1}$$

$$H_{\text{vib}} = \sum_\lambda \frac{p_\lambda^2}{2m} + \frac{m\omega_\lambda^2 x_\lambda^2}{2} = \sum_\lambda \hbar\omega_\lambda a_\lambda^\dagger a_\lambda, \tag{2}$$

$$H_{\text{int}} = H^{(1)} + H^{(2)}, \tag{3}$$

$$H^{(1)} = d^\dagger d \sum_\lambda M_\lambda^{(1)} (a_\lambda + a_\lambda^\dagger) = d^\dagger d \sum_\lambda M_\lambda^{(1)} \sqrt{\frac{2m\omega_\lambda}{\hbar}} x_\lambda, \tag{4}$$

$$H^{(2)} = \frac{1}{2} d^\dagger d \sum_{\lambda\lambda'} (a_\lambda + a_\lambda^\dagger) M_{\lambda\lambda'}^{(2)} (a_{\lambda'} + a_{\lambda'}^\dagger). \tag{5}$$

Here $\omega_\lambda$ denote the unperturbed vibrational frequencies, $x_\lambda = \sqrt{\frac{\hbar}{2m\omega_\lambda}}(a_\lambda + a_\lambda^\dagger)$ is the normal mode coordinate, and $m$ is an effective mass.

## II. VIBRATIONAL GREEN'S FUNCTION

We introduce two related vibrational retarded Green's functions (GFs)

$$D_{x,\lambda\lambda'}^r(t-t') = \frac{1}{i\hbar}\theta(t-t')\langle [x_\lambda(t), x_{\lambda'}(t')]\rangle \tag{6}$$

and

$$D_{\lambda\lambda'}^r(t-t') = -i\theta(t-t')\langle [A_\lambda(t), A_{\lambda'}(t')]\rangle, \tag{7}$$

where $A_\lambda = a_\lambda + a_\lambda^\dagger$ and the latter GF is the one considered in the main part of the paper. The two GFs are related simply by $D_{x,\lambda\lambda'}^r(t) = D_{\lambda\lambda'}^r(t)/2m\sqrt{\omega_\lambda \omega_\lambda'}$.

### A. Vibrational occupation and frequency renormalization

In this section we discuss the interpretation of the vibrational lesser GF and pinpoints its relation to the occupation of the renormalized vibration as well as problems encountered in the published literature. The main conclusion following our considerations is that for weak el-vib coupling, a rate-equation approach for the occupation of the renormalized vibration is essentially exact and can be recovered via a consistent renormalization scheme for the vibrational GF. For simplicity, we here restrict the discussion to the case of a single vibrational mode coupled linearly to the level occupation.

Using the equations of motion for the oscillator coordinates $\dot{x}(t) = \frac{1}{i\hbar}[x,H](t) = p(t)/m$, $\dot{p}(t) = \frac{1}{i\hbar}[p,H](t) = -m\omega_0^2 x(t) - iM\sqrt{\frac{2m\omega_0}{\hbar}}d^\dagger d$ we can derive the following *exact* identities for the retarded GF,

$$\lim_{t' \to t-} D_x^r(t,t') = 0, \tag{8}$$

$$\lim_{t' \to t-} \frac{\partial}{\partial t} D_x^r(t,t') = \frac{1}{i\hbar}\langle[\dot{x}(t),x(t)]\rangle = -\frac{1}{m}. \tag{9}$$

These identities are exact also for *interacting* GFs governed by the full Hamiltonian $H$ (actually, they hold not only for the considered occupation-coordinate coupling but for *any* interaction term as long as it does not contain the momentum operator $p$). Using the above relation between $D_x^r(t)$ and $D^r(t)$ together with the transformation to frequency domain, we derive the following *exact sum rules* for the vibrational spectral function $A(\omega) \equiv -2\Im D^r(\omega)$

$$\int_{-\infty}^{\infty} \frac{d\omega}{2\pi} A(\omega) = 0,$$
$$\int_{-\infty}^{\infty} \frac{d\omega}{2\pi} \omega A(\omega) = 2\omega_0. \tag{10}$$

Once more, these identities hold generally also for interacting GFs. Note, that Eq. (10) contains the *bare frequency* $\omega_0$ regardless of the interaction and consequent renormalization of the value of the frequency. This has important implications for the structure of the interacting GFs as we will see shortly. For the free vibrational GF $D_0^r(\omega) = \frac{1}{\omega-\omega_0+i0^+} - \frac{1}{\omega+\omega_0+i0^+} = \frac{2\omega_0}{(\omega+i0^+)^2-\omega_0^2}$ we get the spectral function $A(\omega) = 2\pi[\delta(\omega-\omega_0)-\delta(\omega+\omega_0)]$ satisfying trivially the above sum rules. When the interaction is included via the polarization operator $\Pi^r(\omega)$ (vibrational self-energy) in the Dyson equation $D^r(\omega)^{-1} = D_0^r(\omega)^{-1} - \Pi^r(\omega)$ one gets

$$D^r(\omega) = \frac{2\omega_0}{\omega^2 - \omega_0^2 - 2\omega_0\Re\Pi^r(\omega) - 2i\omega_0\Im\Pi^r(\omega)} = \frac{\omega_0}{\tilde{\Omega}(\omega)} \cdot \frac{2\tilde{\Omega}(\omega)}{\omega^2 - \tilde{\Omega}^2(\omega) + i\tilde{\Omega}(\omega)\gamma(\omega)}, \tag{11}$$

with the renormalized frequency (so far being a function of the measurement frequency $\omega$) $\tilde{\Omega}^2(\omega) \equiv \omega_0^2 + 2\omega_0\Re\Pi^r(\omega)$ and the damping function $\gamma(\omega) \equiv -2\Im\Pi^r(\omega)\omega_0/\tilde{\Omega}(\omega)$.

For weak coupling when the linewidth is much smaller than the renormalized frequency $\gamma(\omega) \ll \tilde{\Omega}(\omega)$, we can approximate Eq. (11) in the vicinity of the renormalized frequency $\tilde{\omega}_0$ by (kinetic approximation)

$$D^r(\omega) = \frac{2\omega_0}{\omega^2 - \tilde{\omega}_0^2 + i\tilde{\omega}_0\gamma} \approx \frac{\omega_0}{\tilde{\omega}_0} \times \left[\frac{1}{\omega - \tilde{\omega}_0 + i\gamma/2} - \frac{1}{\omega + \tilde{\omega}_0 + i\gamma/2}\right], \tag{12}$$

where $\tilde{\omega}_0$ is determined by the solution of the equation $\tilde{\omega}_0^2 = \tilde{\Omega}_0^2(\tilde{\omega}_0) = \omega_0^2 + 2\omega_0\Re\Pi^r(\tilde{\omega}_0) \approx \omega_0^2 + 2\omega_0\Re\Pi^r(\omega_0)$ and $\gamma \equiv \gamma(\tilde{\omega}_0) = -2\Im\Pi^r(\tilde{\omega}_0)\omega_0/\tilde{\omega}_0$. The corresponding spectral function is given by two Lorentzians of width $\gamma$ and centered at the renormalized frequency $\pm\tilde{\omega}_0$. It should here be noted that the effect of interactions is apart from the frequency renormalization and appearance of finite linewidth $\gamma$ also the multiplicative factor in front of the square bracket. This factor corresponds to the *wave-function renormalization* in the quantum-field-theoretic language and is necessary for the satisfaction of the sum rules (10). In order to emphasize its origin in the vibrational case considered here, we shall refer to it as the *mode-renormalization* factor.

Next, we turn to the vibrational lesser Green's function $D^<(\omega)$. It is determined by the Keldysh equation $D^<(\omega) = D^r(\omega)\Pi^<(\omega)D^a(\omega)$ giving explicitly

$$D^<(\omega) = \frac{4\omega_0^2\Pi^<(\omega)}{[\omega^2 - \tilde{\Omega}^2(\omega)]^2 + 4\omega_0^2|\Im\Pi^r(\omega)|^2} = -A(\omega)\frac{\Pi^<(\omega)}{2\Im\Pi^r(\omega)}. \tag{13}$$

In the kinetic approximation (12) we can reformulate the above Eq. (13) as

$$D^<(\omega) \approx -i\frac{\omega_0}{\tilde{\omega}_0} \times \left[\frac{\gamma}{(\omega-\tilde{\omega}_0)^2 + (\gamma/2)^2}N(\tilde{\omega}_0) + \frac{\gamma}{(\omega+\tilde{\omega}_0)^2 + (\gamma/2)^2}(N(\tilde{\omega}_0)+1)\right], \tag{14}$$

with $N(\tilde{\omega}_0) \equiv \frac{i\Pi^<(\tilde{\omega}_0)}{2|\Im\Pi^r(\tilde{\omega}_0)|}$ defining the mean nonequilibrium occupation identical to the rate-equation expression [Eq. (6) of the main paper]. When $D^<(\omega)$ in Eq. (14) is integrated over the frequency we get

$$i\int \frac{d\omega}{2\pi} D^<(\omega) = \frac{\omega_0}{\tilde{\omega}_0}(2N(\tilde{\omega}_0)+1). \tag{15}$$

Notice that the integral is *not* simply yielding the expected quantity $\langle 2a^\dagger a + 1\rangle \stackrel{?}{=} 2N(\tilde{\omega}_0) + 1$ but the *mode-renormalization* factor is involved. This together with the omission of the frequency renormalization is behind the reported discrepancies between the NEGF calculations and rate-equation predictions in Ref. 1 as we show in detail below.

From our analysis above, we conclude that the physically relevant occupation factor of the renormalized vibration (i.e., the occupation corresponding to, e.g., the experimental Raman temperature) should be identified from the ratio of lesser and retarded self-energies around the renormalized-frequency peak as in Eq. (14). For a renormalized vibration the integral of the lesser GF has *per se* no real physical meaning. The reason for this is that the lesser GF contains the sum $(a+a^\dagger) = \sqrt{\frac{2m\omega_0}{\hbar}} x_0$ related to the physical normal mode coordinate $x_0$ via the frequency-dependent prefactor. However, the bare frequency $\omega_0$ *is not* the physically relevant true dynamical frequency of the renormalized oscillator. So instead of studying the GFs defined via $a$'s with $\omega_0$ we should redefine *physical fields* $\tilde{a}$, $\tilde{a}^\dagger$ related to $x_0$ via the renormalized frequency $\tilde{\omega}_0$. Then $\tilde{D}/2m\tilde{\omega}_0 = D_x = D/2m\omega_0$ relating the GF in terms of renormalized fields $\tilde{D}$ to the bare one considered everywhere above so far implying $\tilde{D} = \frac{\tilde{\omega}_0}{\omega_0} D$ and consequently

$$i \int \frac{d\omega}{2\pi} \tilde{D}^<(\omega) = 2N(\tilde{\omega}_0) + 1 \tag{16}$$

as expected.

*Comparison with results of* D. F. Urban *et al.,* Phys. Rev. B **82**, 121414(R) (2010)

We conclude this part by the direct comparison of our findings with those of Ref. 1. There, integral (15) is identified directly with the occupation number, i.e. $\langle n \rangle = -\frac{1}{2} + \frac{i}{4\pi} \int d\omega D^<(\omega)$ (Ref. 1, p. 3). Rate equation formula gives $N(\Omega) = \frac{V-\Omega}{4\Omega}$ for $V > \Omega$, $T = 0$, and zero external damping $\eta$ (cf. Eq. (10) in Ref. 1). Inserting the right hand-side of our Eq. (15) in the expression for $\langle n \rangle$ above results in $\langle n \rangle = \frac{\omega_0}{4\tilde{\omega}_0} + (\frac{\omega_0}{\tilde{\omega}_0})^2 \frac{V}{4\omega_0} - \frac{1}{2}$. Frequency renormalization in the leading order reads in quantities of Ref. 1 (cf. their Eqs. (7)-(9)) $\Re\Pi^r(\tilde{\omega}_0) \approx -\frac{\lambda^2}{\pi\Gamma}$ which implies $\tilde{\omega}_0 \approx \omega_0 \sqrt{1 - \frac{2\lambda^2}{\pi\Gamma\omega_0}}$. When plugged into the above expression for $\langle n \rangle$ we obtain for the parameters of Fig. 2(a) $V = 5\omega_0$, $\omega_0/\Gamma = 0.01$ the following results $\langle n \rangle \doteq 1.08$ ($\tilde{\omega}_0 \doteq 0.97\omega_0$) for $\lambda/\Gamma = 0.03$ and $\langle n \rangle \doteq 1.26$ ($\tilde{\omega}_0 \doteq 0.92\omega_0$) for $\lambda/\Gamma = 0.05$ which perfectly explain the asymptotic values for zero external damping $\eta \to 0$ shown in Fig. 2(a) of Ref. 1. It should be noticed that the expected rate equation result also does not coincide with the value $N = 1$ mentioned in Ref. 1 where the frequency renormalization was completely omitted. Instead the expected values are $N(\tilde{\omega}_0) \doteq 1.04, 1.11$ for $\lambda/\Gamma = 0.03, 0.05$, respectively. These values are still different form those calculated for $\langle n \rangle$ due to the neglect of the mode renormalization factor in Ref. 1.

### B. Self-energy for the quadratic el-vib interaction

In the following we derive an expression for the lowest-order self-energy for the quadratic el-vib interaction in Eq. (5). Since the diagrammatic rules for nonequilibrium theory is structurally equivalent to equilibrium theory, it is sufficient to consider the time-ordered Green's function at $T = 0^2$.

The time-ordered vibrational GF is defined as

$$D_{\lambda\lambda'}(t,t') = -i\langle T A_\lambda(t) A_{\lambda'}(t')\rangle, \tag{17}$$

where $T$ is the time-ordering operator.

In the interaction picture the time evolution of the operators is governed by the $S$-matrix

$$S(t,t') = T \exp\left[-i\int_{t'}^t dt_1 H_{\text{int}}(t_1)\right] \approx 1 - i\int_{t'}^t dt_1 T[H_{\text{int}}(t_1)], \tag{18}$$

with the last equality giving the lowest-order expansion in the interacting part of the Hamiltonian $H_{\text{int}}$ (in our case the el-vib interaction). The time-ordered GF can be written in terms of the $S$-matrix as

$$D_{\lambda\lambda'}(t,t') = -i\frac{\langle T [A_\lambda(t) A_{\lambda'}(t') S(-\infty,\infty)]\rangle_0}{\langle S(-\infty,\infty)\rangle_0} = D^{\text{con}}_{\lambda\lambda'}(t,t') \tag{19}$$

where $\langle \cdot \rangle_0$ denotes the expectation value with respect to the noninteracting part of the Hamiltonian and $D_{\lambda\lambda'}^{\text{con}}(t,t')$ is the sum of connected diagrams in the numerator[3].

Considering now the quadratic el-vib interaction in Eq. (5) as the interacting part of the Hamiltonian, we insert the lowest-order expansion of the $S$-matrix into the numerator of Eq. (19) and identify the connected diagrams. Using Wick's theorem on the product of time-ordered operators we get

$$\langle T[A_\lambda(t)A_{\lambda'}(t')S(-\infty,\infty)]\rangle_0$$

$$= \langle TA_\lambda(t)A_{\lambda'}(t')\rangle_0 - \frac{i}{2}\sum_{\lambda_1\lambda_2} M_{\lambda_1\lambda_2}^{(2)} \int dt_1 \, \langle TA_\lambda(t)A_{\lambda'}(t')A_{\lambda_1}(t_1)A_{\lambda_2}(t_1)\rangle_0 \langle Td^\dagger(t_1^+)d(t_1)\rangle_0$$

$$= \langle TA_\lambda(t)A_{\lambda'}(t')\rangle_0 - \frac{i}{2}\sum_{\lambda_1\lambda_2} M_{\lambda_1\lambda_2}^{(2)} \int dt_1 \Bigg[ \underbrace{\langle TA_\lambda(t)A_{\lambda'}(t')\rangle_0 \langle TA_{\lambda_1}(t_1)A_{\lambda_2}(t_1)\rangle_0}_{\text{disconnected}}$$

$$+ \underbrace{\langle TA_\lambda(t)A_{\lambda_1}(t_1)\rangle_0 \langle TA_{\lambda_2}(t_1)A_{\lambda'}(t')\rangle_0}_{\delta_{\lambda\lambda_1}\delta_{\lambda'\lambda_2}} + \underbrace{\langle TA_\lambda(t)A_{\lambda_2}(t_1)\rangle_0 \langle TA_{\lambda_1}(t_1)A_{\lambda'}(t')\rangle_0}_{\delta_{\lambda\lambda_2}\delta_{\lambda'\lambda_1}} \Bigg] \langle Td^\dagger(t_1^+)d(t_1)\rangle_0. \quad (20)$$

For the GF we thus have

$$D_{\lambda\lambda'}(t,t') = D_{\lambda\lambda'}^0(t,t') + \int dt_1 \, D_{\lambda\lambda}^0(t,t_1)\left[-iG(t_1,t_1^+)M_{\lambda\lambda'}^{(2)}\right]D_{\lambda'\lambda'}^0(t_1,t')$$

$$\equiv D_{\lambda\lambda'}^0(t,t') + \int dt_1 \int dt_2 \, D_{\lambda\lambda}^0(t,t_1)\Pi_{\lambda\lambda'}^{(2)}(t_1,t_2)D_{\lambda'\lambda'}^0(t_2,t'), \quad (21)$$

where $D_{\lambda\lambda'}^0 = \delta_{\lambda\lambda'}D_{\lambda\lambda}^0$ is the noninteracting GF and $G(t,t') = -i\langle Td(t)d^\dagger(t')\rangle$ is the time-ordered electronic GF. In the last equation, we have identified the irreducible lowest-order self-energy for the quadratic el-vib interaction as

$$\Pi_{\lambda\lambda'}^{(2)}(t,t') = -iM_{\lambda\lambda'}^{(2)}\delta(t-t')G(t,t^+) \quad (22)$$

with the retarded and lesser versions given by

$$\Pi_{\lambda\lambda'}^{(2),r}(t-t') = -iM_{\lambda\lambda'}^{(2)}\delta(t-t')G^<(t-t'=0) \quad (23)$$

$$\Pi_{\lambda\lambda'}^{(2),<}(t-t') = 0. \quad (24)$$

In frequency domain they become

$$\Pi_{\lambda\lambda'}^{(2),r}(\omega) = -iM_{\lambda\lambda'}^{(2)}\int \frac{d\varepsilon}{2\pi} \, G^<(\varepsilon) \quad (25)$$

$$\Pi_{\lambda\lambda'}^{(2),<}(\omega) = 0. \quad (26)$$

## III. FIRST-PRINCIPLES CALCULATIONS

In this section we present the results of our first-principles calculations of the vibrational Stark shifts and el-vib interactions for the OPV3 molecule. All calculations were performed with the GPAW electronic structure package[4,5] using the finite-difference method for the calculation of the molecular vibrations and el-vib interactions[6,7]. The calculations have been carried out on the isolated amine-terminated OPV3 molecule, thus neglecting direct effects of lead electrodes on the calculated quantities.

### A. Vibrational Stark shifts

In order to clarify the role of the vibrational Stark effect[8] in the shifts of the Raman lines observed in Ref. 9, we have calculated the vibrational frequencies of the OPV3 molecule in the presence of an electric field along the backbone axis of the molecule. The vibrational Stark shifts have been obtained as the change $\Delta\omega_\lambda$ in the vibrational frequencies in the presence of a potential drop $V$ along the longest axis of the computational unit cell of the isolated molecule ($L \sim 3$ nm) corresponding to a constant electric field $E = V/L$ over the extend of the molecule.

Figure 1 shows the resulting frequency shifts $\Delta\omega_\lambda$ for different values of the potential drop over the molecule. For all modes with $\omega_\lambda \gtrsim 60$ meV no significant effect of the electric field is observed. We also note that the Stark effect is static and does not produce linewidth broadening. We therefore conclude that the Stark effect is not a likely source of the frequency shifts and linewidth broadening observed in Ref. 9.

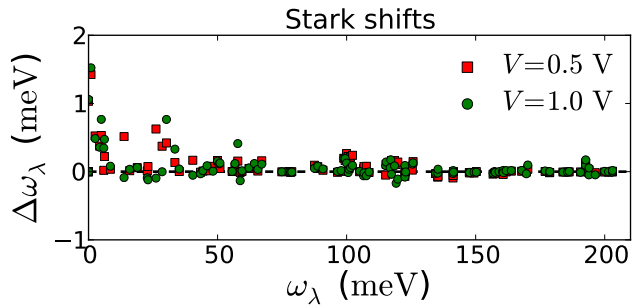

FIG. 1: (Color online) Calculated vibrational Stark shifts in the amine-terminated OPV3 molecule (see text for details).

### B. El-vib interactions

The el-vib interactions in Eqs. (4) and (5) have been obtained following the approach outlined in Ref. 6. Within this approach, the el-vib interactions for the molecular orbital $\psi_0$ of the isolated molecule can be expressed as

$$M_\lambda^{(1)} = \sqrt{\frac{\hbar}{2M\omega_\lambda}} \langle \psi_0 | \hat{\mathbf{e}}_\lambda \cdot \nabla H | \psi_0 \rangle \tag{27}$$

and

$$M_{\lambda\lambda'}^{(2)} = \frac{\hbar}{2M\sqrt{\omega_\lambda \omega_\lambda'}} \langle \psi_0 | \hat{\mathbf{e}}_\lambda \cdot \nabla\nabla H \cdot \hat{\mathbf{e}}_{\lambda'} | \psi_0 \rangle, \tag{28}$$

for the linear and quadratic el-vib interaction, respectively. Here, $M$ is an appropriately defined effective mass, $\hat{\mathbf{e}}_\lambda$ is the mass-scaled vibrational normal mode vector normalized according to $\sum_\alpha (M_\alpha/M)|\hat{\mathbf{e}}_{\lambda\alpha}|^2 = 1$ where $M_\alpha$ is the mass of atom $\alpha$, and the gradient of the Hamiltonian $H$ is with respect to atomic displacements. The expressions (27) and (28) can be understood as the first-order change in the orbital energy $\varepsilon_0$ due to the atomic displacements associated with the vibrational normal modes.

The calculation of the el-vib couplings in Eqs. (27) and (28) have been performed using the finite-difference method[6]. The implementation for the calculation of the linear el-vib interaction in the GPAW package is outlined in Ref. 7. For the quadratic el-vib interaction, the couplings have been obtained as the second-order derivative of the orbital energy $\varepsilon_0$ with respect to a unit displacements $u_\lambda$ in the normal mode directions,

$$M_{\lambda\lambda'}^{(2)} = \frac{\hbar}{2M\sqrt{\omega_\lambda \omega_{\lambda'}}} \frac{\partial^2 \varepsilon_0}{\partial u_\lambda \partial u_{\lambda'}}. \tag{29}$$

The calculated couplings for the OPV3 molecule are summarized in Fig. 2 which shows the absolute value of the linear (left) and diagonal elements of the quadratic (right) el-vib couplings. For both the linear and quadratic interaction, the couplings to the HOMO and LUMO orbitals are highly mode sensitive. The linear couplings are rather small ($|M_\lambda^{(1)}| < 5$ meV) for many of the modes and therefore hardly have any impact on the corresponding spectral lines (see below). The quadratic couplings are negative ($M_{\lambda\lambda}^{(2)} < 0$) for the majority of modes. This corresponds to frequency *softening* and *hardening* for charging of the LUMO and HOMO, respectively (charging of a filled HOMO corresponds to electron removal, implying that the associated frequency shift has the opposite sign of the quadratic coupling). These findings are in good agreement with a recent study of charging-induced frequency shifts in molecular junctions[10].

### IV. OPV3 SPECTRAL FUNCTION

In this section we present the details of the calculation of the vibrational spectral function for the OPV3 junction, as well as results for both HOMO and LUMO dominated transport.

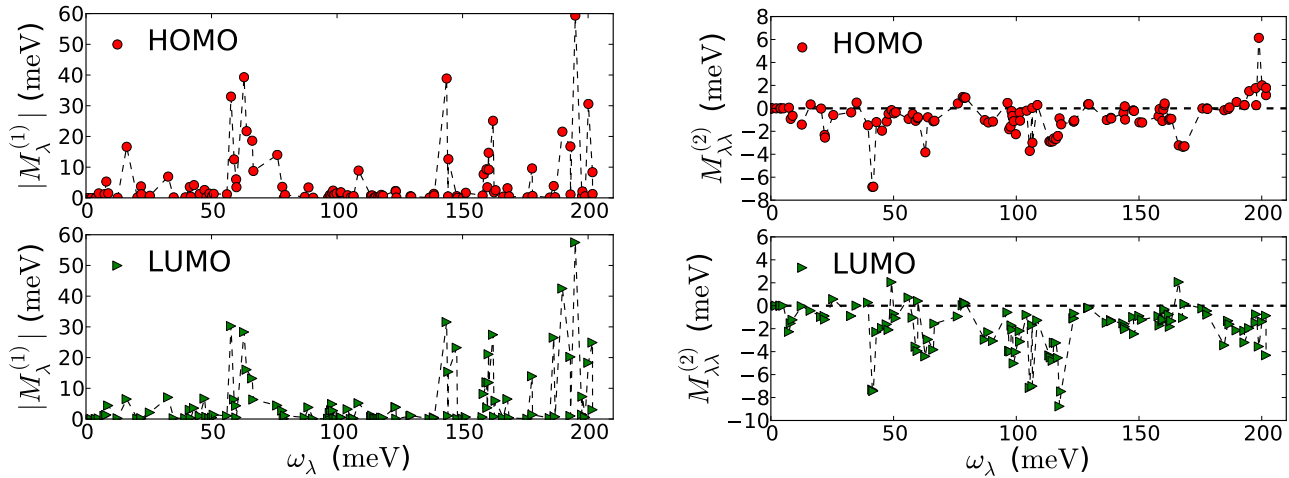

FIG. 2: (Color online) Calculated el-vib interactions in the OPV3 molecule as a function of the vibrational frequencies. (Left) Absolute value of the linear el-vib interaction $M_\lambda^{(1)}$. (Right) Diagonal elements $M_{\lambda\lambda}^{(2)}$ of the quadratic el-vib interaction.

### A. Calculational details

The spectral function of the OPV3 junction is obtained from the full non-diagonal GF (cf. Eq. (1) of the main paper)

$$\mathbf{D}^r(\omega) = \left[[\mathbf{D}_0(\omega)]^{-1} - \mathbf{\Pi}^r(\omega)\right]^{-1}, \quad (30)$$

where carrier-induced mode-mode couplings are accounted for via the off-diagonal elements of the electronic part $\Pi^{\text{el}}$ of the total self-energy $\Pi = \Pi^{\text{el}} + \Pi^{\text{ph}}$. The bare GF $\mathbf{D}_0^r(\omega)$ is here given by that of the isolated neutral molecule with frequencies $\omega_\lambda$.

Following our footnote in Ref. 29 of the main paper, effects already included in the frequencies $\omega_\lambda$ of the isolated molecule have to be removed from the self-energy in Eq. (30) in order to avoid double counting. Screening effects due to eh-pair excitations are *not* included in the frequencies $\omega_\lambda$ of the isolated molecule (no lead-induced broadening of the electronic levels) and the self-energy of the linear el-vib interaction should therefore not be corrected. On the other hand, charging effects described by the self-energy of the quadratic el-vib interaction, must be measured relative to the level occupations in the neutral molecule. The self-energy is therefore corrected according to

$$\Pi^{\text{el}}(\omega) \to \Pi^{\text{el}}(\omega) - \Pi^{\text{el}}_{\text{mol}}(\omega), \quad (31)$$

where $\Pi^{\text{el}}_{\text{mol}}$ is the self-energy correction of the neutral isolated molecule given solely by the quadratic self-energy correction as

$$\Pi^{\text{el}}_{\text{mol},\lambda\lambda'}(\omega) = \Pi^{(2)}_{\lambda\lambda'}(\omega) = \langle n_i \rangle M^{(2)}_{\lambda\lambda'} = \begin{cases} 2M^{(2)}_{\lambda\lambda'}, & i = \text{HOMO} \\ 0, & i = \text{LUMO}. \end{cases} \quad (32)$$

Here the factor of 2 accounts for the spin degeneracy of the HOMO level.

### B. Results

The calculated spectral functions for HOMO and LUMO dominated transport are shown in Figs. 3 and 4, respectively. Due to the large variation in the el-vib couplings to the different vibrational modes, carrier-induced shift and broadening of the spectral lines is highly mode dependent. Overall, the quadratic el-vib interaction gives rise frequency *softening* and *hardening* for LUMO and HOMO dominated transport, respectively. However, for the high-energy modes with positive quadratic coupling to the HOMO (see Fig. 2) this does not hold. On top of the frequency renormalization given by the quadratic el-vib interaction, the modes with the strongest the linear el-vib

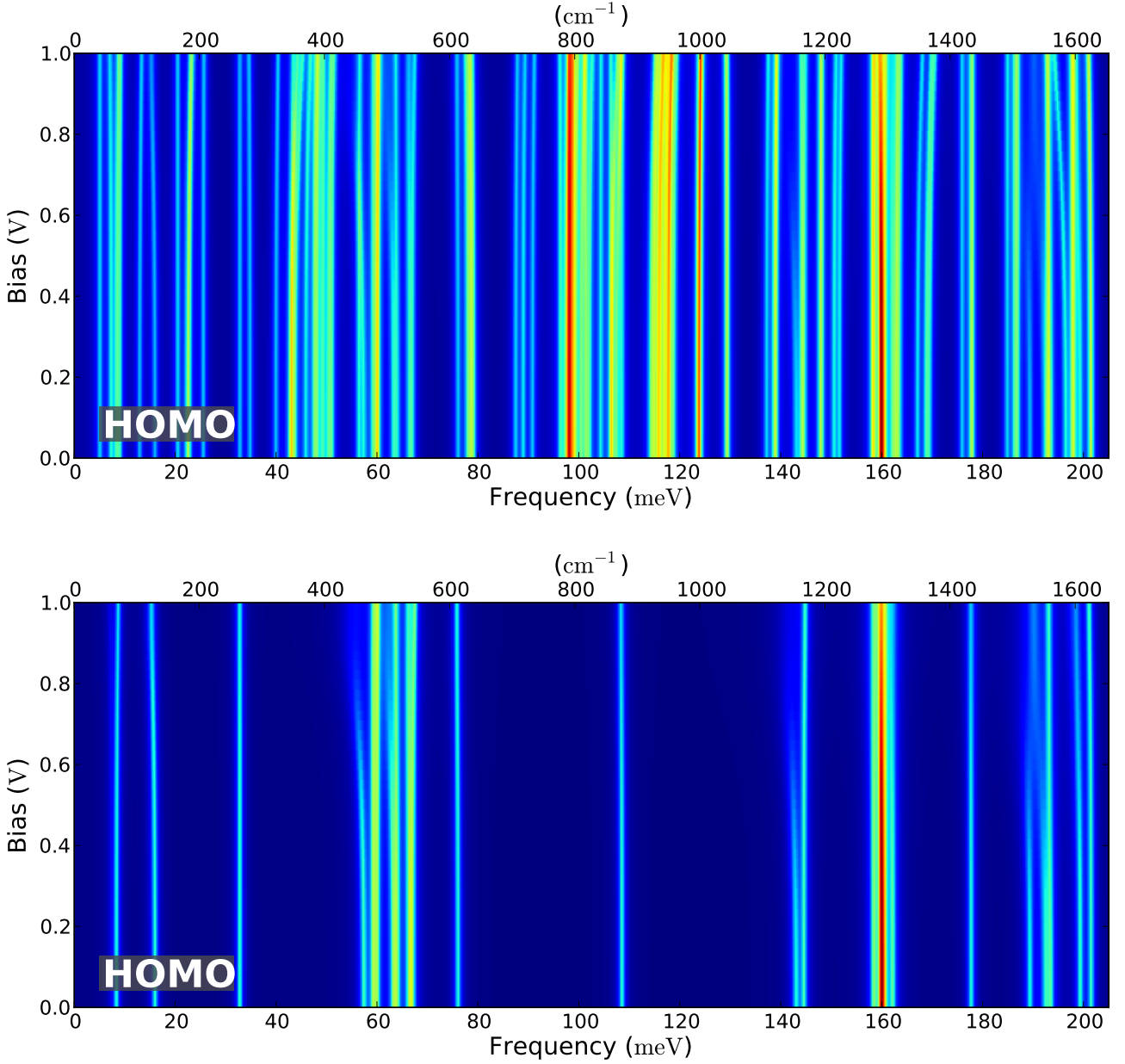

FIG. 3: (Color online) Vibrational spectral function $A(\omega)$ as a function of bias voltage for HOMO-dominated transport in the OPV3 junction (top: full spectral function; bottom: projection onto vibrations with linear coupling $|M_\lambda^{(1)}| > 5$ meV). Except for the level position, the junction parameters are the same as those used in Fig. 3 of the main paper, i.e. $\varepsilon_0 = -0.5$ eV, $\mu_{L/R} = \pm V/2$, $\Gamma_{L/R} = 0.2$ eV, and $\gamma_{\mathrm{ph}} = 1$ meV.

interaction experience additional renormalization and pronounced spectral broadening with increasing bias voltage. The frequency renormalization originating from the linear el-vib interaction corresponds to frequency softening, and thus compensates for (adds to) the renormalization due to the quadratic interaction for HOMO (LUMO) dominated transport. For most of the modes with a significant linear el-vib coupling, the two effects are of the same order of magnitude. As mentioned in the main paper, the relative magnitude of the two effects is sensitive to junction parameters such as the level alignment and lead-induced level broadening. A situation where one effect dominates the other may therefore be realized with a different choice of junction parameters. As a final remark, we note that some of the closely lying spectral lines show complex bias dependencies at high bias which is an indication of carrier-induced mode-mode couplings.

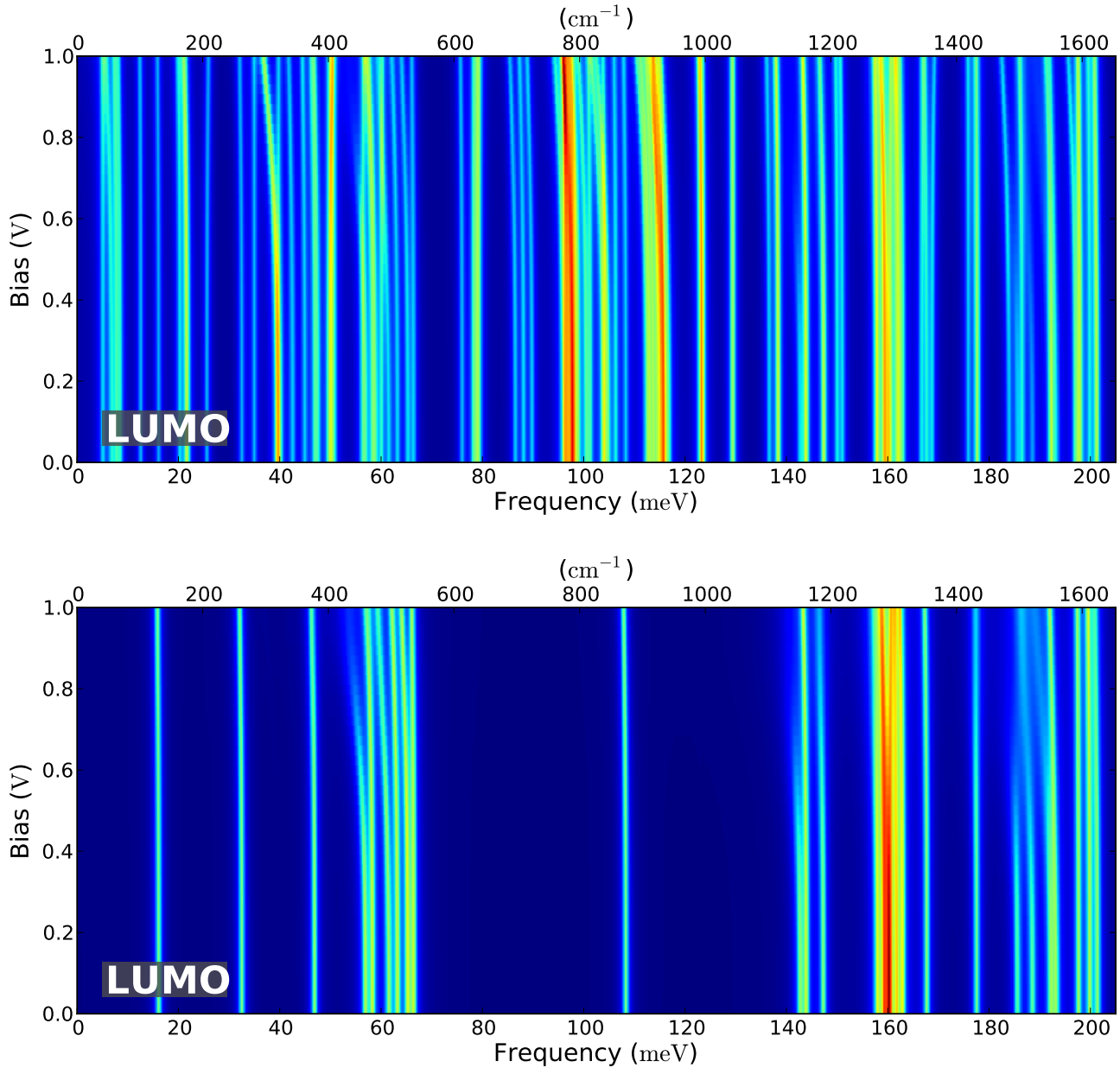

FIG. 4: (Color online) Vibrational spectral function $A(\omega)$ as a function of bias voltage for LUMO-dominated transport in the OPV3 junction (top: full spectral function; bottom: projection onto vibrations with linear coupling $|M_\lambda^{(1)}| > 5$ meV). The junction parameters are the same as those used in Fig. 3 of the main paper, i.e. $\varepsilon_0 = 0.5$ eV, $\mu_{L/R} = \pm V/2$, $\Gamma_{L/R} = 0.2$ eV, and $\gamma_{\rm ph} = 1$ meV.




\begin{thebibliography}{10}
\bibitem{footnote4}%
  \BibitemOpen
  \bibinfo {note} {In spite of the non-conservative nature of the first Born
  approximation applied here, it is expected to give a quantitative description
  of vibrational heating in the regime of weak el-vib
  interaction~\cite{Wang:Coupled}.}%
  \bibAnnoteFile{Stop}{footnote4}%
\bibitem{footnote6}%
  \BibitemOpen
  \bibinfo {note} {Our choice of parameters for the OPV3 junction corresponds
  to a zero-bias conductance of $G\sim 0.1\,G_0$ ($G_0=2e^2/h$) which is
  slightly larger than the conductance values reported in
  Ref.~\onlinecite{Natelson:Heating}.}%
  \bibAnnoteFile{Stop}{footnote6}%
\bibitem{footnote3}%
  \BibitemOpen
  \bibinfo {note} {Note that the cross section for Raman scattering is mode
  dependent implying that not all spectral lines show up in Raman
  spectroscopy.}%
  \bibAnnoteFile{Stop}{footnote3}%
\end{thebibliography}
\end{document}